\documentclass[%
 reprint,
 amsmath,amssymb,
 aps,
floatfix,
]{revtex4-2}

\usepackage{graphicx}
\usepackage{dcolumn}
\usepackage{bm}
\usepackage{textcomp}
\usepackage{gensymb}
\usepackage{amsmath}
\usepackage{booktabs}  
\usepackage{multirow}
\usepackage{makecell}  

\usepackage{xcolor}
\usepackage{hyperref}

\begin{document}

\preprint{APS/123-QED}

\title{Quantum Keyless Private Communication under intense background noise}


\author{Pedro Neto Mendes$^{1,2}$}
\author{Davide Rusca$^{3,4,5}$}
\author{Hugo Zbinden$^{3,4,5}$}
\author{Emmanuel Zambrini Cruzeiro$^{1,2}$}
\email{emmanuel.cruzeiro@lx.it.pt}

\affiliation{$^1$Instituto de Telecomunicações, Av. Rovisco Pais, 1049-001, Lisbon, Portugal}
\affiliation{$^2$Departamento de Engenharia Electrotécnica e de Computadores, Instituto Superior Técnico, Av. Rovisco Pais, 1049-001, Lisbon, Portugal}
\affiliation{$^3$Vigo Quantum Communication Center, University of Vigo, Vigo E-36310, Spain}
\affiliation{$^4$Escuela de Ingeniería de Telecomunicación, Department of Signal Theory and Communications, University of Vigo, Vigo E-36310, Spain}
\affiliation{$^5$AtlanTTic Research Center, University of Vigo, Vigo E-36310, Spain}


\date{\today}

\begin{abstract}
Quantum key distribution relies on quantum mechanics to securely distribute cryptographic keys, offering security but necessitating complex infrastructure and significant resources for practical implementation. Quantum keyless private communication ensures information-theoretic security in free-space communication, with simpler setups, and without the need for secret keys by leveraging the wiretap channel model. This model requires assumptions about Eve’s ability to interact with the channel, consistent with the framework of physical-layer security. Here, we propose a variant of quantum keyless private communication using polarization encoding and experimentally validate both the original on-off keying method and the polarization-multiplexed approach using time-multiplexed threshold single-photon detectors as photon counting detectors. Our analysis highlights the advantages of polarization-multiplexed schemes for daylight operation. This work paves the way towards practical and scalable quantum communication systems, with potential applications extending to space-based communication.
\end{abstract}

\maketitle


\section{Introduction}\label{sec:intro}

Quantum communication encompasses a wide range of protocols that have the potential to revolutionize our understanding of information transfer. Quantum Key Distribution (QKD) has attracted the most attention among these protocols, namely, it is the first commercial solution for quantum communication. Despite the maturity of QKD, its implementation over long distances represents a significant challenge \cite{boaron2018secure}. For example, space-based QKD, while theoretically feasible, is currently limited to nighttime operation, and while recent studies propose methods to address this limitation, they do not enable practical secret key rates \cite{zheng2025free, cai2024free, ko2018experimental, abasifard2024ideal}. Achieving significantly higher key rates at lower system costs is critical for quantum communication to become globally widespread. Therefore, it is imperative to explore new solutions that demand fewer experimental resources but still guarantee security based on the laws of quantum mechanics.

Quantum Keyless Private Communication (QKPC) \cite{vazquez2021quantum} is based on the wiretap channel model \cite{wyner1975wire, wyner1988capacity2} and provides information-theoretical security without the need for a pre-shared secret key. This method assumes that legitimate users are constrained by state-of-the-art technology, while the eavesdropper is limited by both physical laws and restricted access to the communication channel. As QKPC relies on physical constraints on the eavesdropper’s access to the channel, rather than on computational assumptions, it is inherently immune to “store-now-decrypt-later” attacks.

The original QKPC protocol \cite{vazquez2021quantum}, OOK-QKPC herein, employs On-Off Keying (OOK) with coherent states to demonstrate positive private capacity over a classical-quantum wiretap channel. This approach has been shown to achieve much higher rates compared to QKD and is significantly less sensitive to noise and signal dynamics \cite{mendes2024optical}. Recently, in \cite{vazquez2024quantum}, it has been found that it is possible to ensure positive secrecy capacity even when the eavesdropper captures a large part of the photon energy detected by Bob, at the cost of making the wiretap channel asymmetric both for Bob and Eve. This is achieved by adding artificial noise to the encoder.

In this work, we employ a time-multiplexing detection scheme to approximate photon number resolving (PNR) measurements and show that this improves the robustness to background noise of OOK-QKPC. We then generalize OOK-QKPC by proposing a polarization-multiplexed version of QKPC, PM-QKPC for short. Finally, we show theoretically and experimentally that the combination of the time-multiplexing scheme and PM-QKPC enables QKPC with significant background noise. In Section \ref{sec:concept}, we derive the private capacities for OOK-QKPC and PM-QKPC using PNR measurements. Section \ref{sec:experiment} describes the experimental demonstration of this scheme using a portable setup. The results are presented in Section \ref{sec:Results}, where we also estimate the private capacity under different photon noise regimes. Subsequently, Section \ref{sec:Comparison} compares our protocol's implementation with QKD under daylight conditions. Finally, Section \ref{sec:conclusion} offers conclusions and provides an outlook for future work.

\section{Quantum Keyless Private Communication}
\label{sec:concept}

\subsection{Overview}

QKPC, introduced in \cite{vazquez2021quantum}, is a direct messaging quantum communication protocol with key advantages over other quantum direct messaging protocols like quantum secure direct communication \cite{pan2023free}: 1) It works in a one-way configuration, simplifying implementation; 2) It relies on a discrete-variable encoding using coherent state (e.g., OOK), ensuring experimental simplicity; 3) No two-way classical communication is needed, which is beneficial for satellite communications.

QKPC is based on the classic wiretap model \cite{wyner1975wire}, where Alice sends a message to Bob over a channel eavesdropped by Eve. The goal is to encode the message in a way that maximizes Eve's confusion, preventing her from decoding it. In some free-space links, satellite communications for example, an eavesdropper would need to insert detection apparatus into the transmission path while also satisfying orbital and geometric constraints, and would still face intrusion monitoring (e.g., losses, timing, spatial modes), making undetected interception practically infeasible. This places such scenarios naturally within the framework of physical-layer security, an established approach in classical communication systems \cite{poor2017wireless, endo2016free}, and here proposed as a viable alternative to QKD for satellite-to-ground quantum communication.

Assuming a single-mode free-space quantum bosonic channel, the efficiency is denoted as $\eta$ which incorporates losses arising from beam divergence, pointing errors, atmospheric turbulence and absorption, as well as intrinsic losses within Bob's setup. The channel degradation is described by $\gamma \in (0,1)$, so Eve’s channel efficiency is $\gamma\eta$, lower or equal to Bob's (as required by the wiretap model), with a detailed model of this channel present in \cite{vazquez2021quantum}. Bob is assumed to have a single photon detector with limited efficiency (included in $\eta$) and dark count probability $p_\text{dark}$.

QKPC involves four main steps: 1) Classical encoding of the message; 2) Quantum state preparation; 3) State discrimination by the receiver; 4) Decoding to extract the message. The encoding and decoding constructions, assumed to be public, enhance security by adding confusion for Eve \cite{vazquez2021quantum, vazquez2024quantum}. The PM-QKPC protocol proposed here extends OOK-QKPC by modifying the state preparation and discrimination steps, without altering the encoding or decoding procedures.

\subsection{OOK-QKPC}

We consider a binary channel with input probabilities $q_0$ and $q_1 = 1 - q_0$. Depending on the input bit $x\in \{0,1\}$, the prepared state is either the vacuum or a coherent state, prepared as in Eq. \ref{Eq: OOK states}:

\begin{equation}
\begin{aligned}
    |\psi_{x=0} \rangle = |\psi_0 \rangle &= |0\rangle, \\
    |\psi_{x=1} \rangle = |\psi_1 \rangle &= |\alpha \rangle.
\end{aligned}
\label{Eq: OOK states}
\end{equation}

Bob uses PNR detectors to measure the incoming states. For each state sent, the detector counts the number of arriving photons. To discriminate the states, we use a simple strategy: if the detector registers a number of photons equal or greater than a given threshold, k, we assume the state sent was $|\psi_1 \rangle$ (as this state has a higher average photon number than $|\psi_0 \rangle$), and the bit recorded is 1. If the detector measures a number of photons lower than k, the bit recorded is 0. For $k=1$, no PNR capabilities are necessary, threshold single-photon detectors are sufficient.

Using this strategy, we can characterize the binary communication channel. We assume there are $\Delta$ detector clicks per pulse on average from background photons or/and dark counts, photon noise, and these clicks in the detector follow a Poisson distribution and are independent of the emitter signal. The probability of sending bit 0 and recording bit 0 is given by the probability of sending state $|\psi_0 \rangle$ and measuring fewer than $k$ photons. This probability is given by:

\begin{equation}
    \epsilon_{00} = \sum_{i=0}^{k - 1} e^{-\Delta} \frac{\Delta^i}{i!}.
\end{equation}

The probability of sending bit 1 and recording bit 0 is given by the probability of sending state $|\psi_1 \rangle$ and measuring fewer than $k$ photons. This probability is given by:

\begin{equation}
    \epsilon_{10} = \sum_{i=0}^{k - 1} e^{-(\eta|\alpha|^2 +\Delta)} \frac{(\eta|\alpha|^2 +\Delta)^i}{i!}.
\end{equation}

The binary communication channel is then fully defined as $\epsilon_{01} = 1 - \epsilon_{00}$ and $\epsilon_{11} = 1 - \epsilon_{10}$. 

The probability of a discrimination error for a chosen $k$ is:

\begin{equation}
    \text{QBER} = \frac{\epsilon_{01} + \epsilon_{10}}{2},
    \label{eq: QBER}
\end{equation}

where we consider a uniform input distribution (i.e. $q_0$ equals $q_1$) as we found it to be very close to the optimal value and simplifies the pre- and post-processing.
The mutual information for Bob, $I_B$, is given by:

\begin{equation}
    I_B = h_b\left(\frac{\epsilon_{00} + \epsilon_{10}}{2}\right) - \frac{h_b(\epsilon_{00}) + h_b(\epsilon_{10})}{2},
    \label{eq: Mutual_information_Bob}
\end{equation}

where $h_b$ is the binary entropy function.
Eve is assumed to have a binary symmetric channel with error probability $\epsilon_{\gamma}$, given by the Helstrom bound. We assume Eve only intercepted a fraction, $\gamma$, of the signal Bob received. For the states assumed in Eq. \ref{Eq: OOK states},

\begin{equation}
    \epsilon_{\gamma} = \frac{1}{2}\left(1-\sqrt{1-|\langle\psi_0|\psi_1\rangle|^2}\right),
    \label{eq: Qber_helstrom}
\end{equation}

with

\begin{equation}
    \langle\psi_0|\psi_1\rangle = e^{-\gamma\eta\frac{|\alpha|^2}{2}}.
\end{equation}

For orthogonal states, the overlap is zero, allowing Eve to perfectly discriminate between them without error, rendering the communication entirely insecure.

Eve's mutual information, $I_E$, is then:

\begin{equation}
    I_E = 1 - h_b(\epsilon_{\gamma}),
    \label{eq: Mutual_information_Eve}
\end{equation}

and the private capacity is:

\begin{equation}
    C_p(\gamma) = \text{max}_{|\alpha|^2, k} \left[I_B - I_E\right].
    \label{eq: Private Capacity}
\end{equation}

If counting the number of single photons arriving at the detector is not possible, only a threshold detector is available, the only threshold choice possible will be $k = 1$.

\subsection{PM-QKPC}

We consider again the same binary channel. For the input bit 0 (1), the state $|\psi_0 \rangle$ ($|\psi_1 \rangle$) is prepared. Two general weak coherent states with polarization in the XZ plane were considered (see Appendix \ref{appendix: discrimination}) but it was found that the optimal case corresponds to having two coherent states with approximately the same average number of photons as in Eq. \ref{Eq: Polarization states simplified}.

\begin{equation}
\begin{aligned}
    |\psi_0 \rangle &= |\alpha\rangle_H \otimes |0\rangle_V, \\
    |\psi_1 \rangle &= |\cos\left(\theta\right)\alpha\rangle_H \otimes |\sin\left(\theta\right)\alpha\rangle_V.
\end{aligned}
\label{Eq: Polarization states simplified}
\end{equation}

To maximize the discrimination probability, we perform a projective measurement by rotating the polarization of the states by $\delta$ before the measurement on the Z basis. For the chosen states, $\delta = -\frac{\theta}{2} + \frac{\pi}{4}$ maximizes the number of photons that go to one detector for a given input state while minimizing it for the other input state.
  
Using two PNR detectors, we adopt a majority-click rule: output is 0 if the first detector registers more counts, else it is 1 (ties $\!\rightarrow\!1$, see Appendix \ref{appendix: discrimination}). Other strategies like unambiguous state discrimination, were studied, but this was found to be the best strategy (see Appendix \ref{appendix: USD}).

For an input bit value 0, the probability of correct discrimination is given by the probability that detector 0 clicks more times than detector 1. The probability of a difference of $m$ clicks between both detectors is given by the convolution of each detector distribution conditioned on the input bit:

 \begin{equation}
     P(m|0) = e^{-\left(\eta|\alpha |^2 + 2\Delta\right)} \sum_{l=0}^{\infty} \frac{(\eta|\alpha_{0} |^2 + \Delta)^{l+m}(\eta|\alpha_{1} |^2 + \Delta)^l}{(l+m)!l!},
 \end{equation} 

 \begin{equation}
     P(m|1) = e^{-\left(\eta|\alpha |^2 + 2\Delta\right)} \sum_{l=0}^{\infty} \frac{( \eta|\alpha_{1} |^2 + \Delta)^{l+m}( \eta|\alpha_{0} |^2 + \Delta)^l}{(l+m)!l!}.
 \end{equation} 

Here, $|\alpha_0|^2$ and $|\alpha_1|^2$ denote the average number of photons arriving at detector~0 and detector~1, respectively.

The probability of correct discrimination for input $x=0$, $\epsilon_{00}$ and of making an error for input $x=1$, $\epsilon_{10}$, are given by:

\begin{equation}
\begin{aligned}
   \epsilon_{00} = P(m \geq 0|0) = \sum_{m=0}^{\infty} P(m|0), \\
   \epsilon_{10} = P(m \geq 0|1) = \sum_{m=0}^{\infty} P(m|1).
\end{aligned}
\end{equation}

The channel is again characterized by $\epsilon_{00}$ and $\epsilon_{10}$. The probability of a discrimination error is then given by Eq. \ref{eq: QBER} and the mutual information by Eq. \ref{eq: Mutual_information_Bob}.

An eavesdropper intercepting a fraction $\gamma$ of the signal has a discrimination error given by:

\begin{equation}
    \epsilon_{\gamma} = \frac{1}{2}\left(1-\sqrt{1-|e^{-\gamma|\alpha|^2\left(1-\cos(\theta)\right)}|^2}\right),
    \label{eq: Qber_helstrom_pol}
\end{equation}

and the privacy capacity can then be calculated using Eq. \ref{eq: Private Capacity} and maximizing also over the new parameters of PM-QKPC (polarization angle, relative number of photons between states).

\section{Experimental implementation}
\label{sec:experiment}

\begin{figure}[ht]
  \centering
  \includegraphics[width=0.45\textwidth]{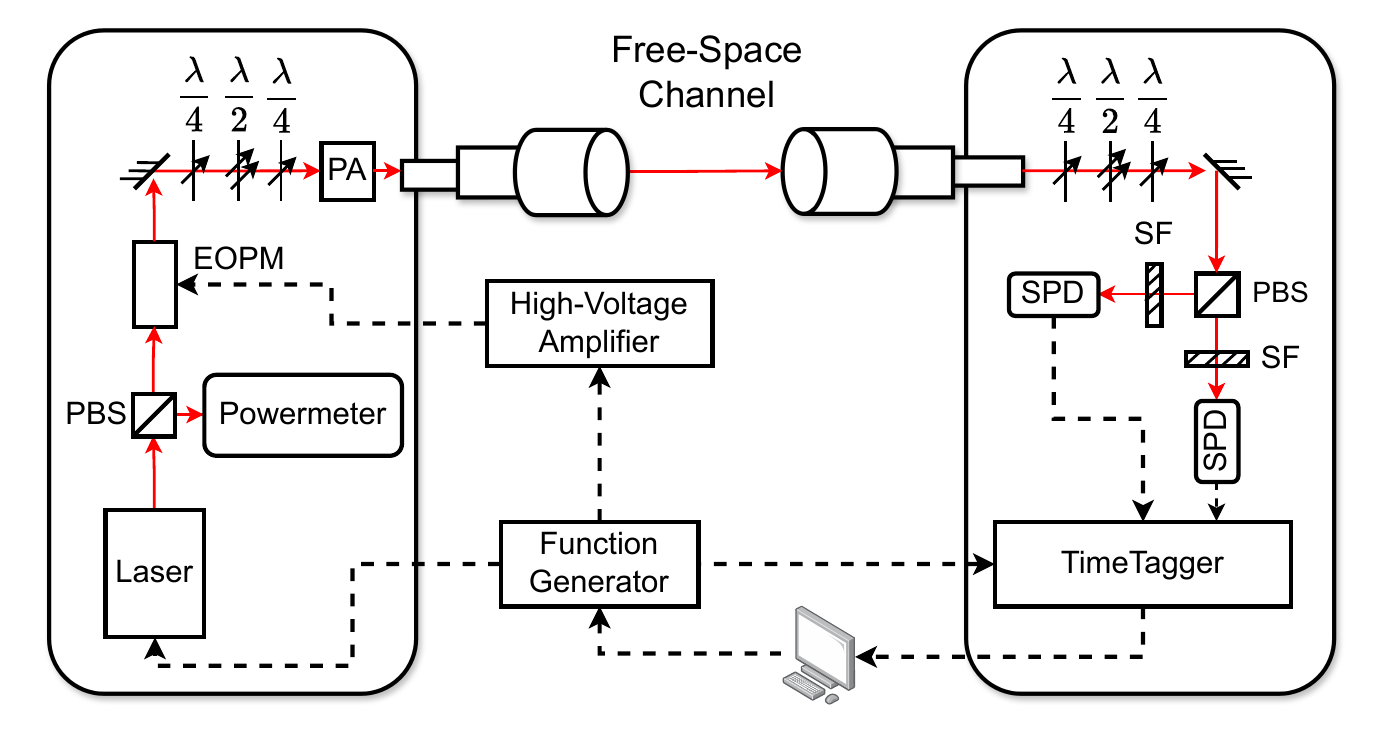} 
  \caption{Experimental setup capable of implementing OOK- and PM-QKPC. PBS: polarizing beam splitter. EOPM: electro-optic polarization modulator. PA: passive attenuator. SPD: single photon detector. $\frac{\lambda}{4}$: quarter-waveplate. $\frac{\lambda}{2}$: half-waveplate. SF: spectral filter. The black dashed arrows are electrical connections.}
  \label{fig:setup}
\end{figure}

The experimental setup capable of implementing both OOK-QKPC and PM-QKPC in a free-space link (3.5 m) is illustrated in Fig. \ref{fig:setup}. The emitter includes a 850 nm laser, Roithner VC850MZ, which is gain-switched using a function generator, AIM-TTI TG330, to generate the pulses. These pulses are sent through a polarizing beam splitter cube (PBS), Thorlabs PBS102, where the vertically polarized component is measured by a power meter, Thorlabs S120VC. The horizontally polarized component is sent through an electro-optic polarization modulator (EOPM), Thorlabs EO-AM-NR-C1, controlled by the function generator and connected to a high voltage amplifier (HVA), Thorlabs HVA200, and a polarization controller (PC), which consists of two quarter-wave plates and one half-wave plate, Thorlabs WPQ05M-850 and WPH05M-850. These components rotate the polarization of each pulse to the desired state. The pulse, now with a well-defined polarization, is attenuated using a combination of optical filters, Thorlabs NE510B-B, to a specific average number of photons, creating a polarized weak coherent pulse.

Both the transmitter and receiver use a beam expander, Thorlabs GBE03-B, with a magnification factor of 3 times. On the receiver side, after the beam expander, a PC corrects any polarization rotations induced by the channel and rotates the states to the polarization which maximizes the discrimination probability. A PBS is then used to separate the photons into two single-photon threshold detectors, Thorlabs SPDMH3 (Dark count $\approx 70$ Hz; Photon detection efficiency $\approx$ 50\%; Dead time $\approx 45$ ns), which are connected to a time tagger, Swabian Time Tagger 20, that registers the detectors' clicks. A spectral filter centered at 850 nm with 10 nm width, Thorlabs FBH05850-10, is used to block background photons. We monitor the pulse generation times by sending the trigger of the function generator to the time-tagger, which works for our proof of principle demonstration. To experiment with larger distances, different synchronization techniques could be considered, either using classical signals \cite{berra2023synchronization} or the quantum signal itself \cite{calderaro2020fast}.

Using the time tagger's data, we can count the number of events in the detectors and estimate the number of photons arriving in each pulse, i.e. we can implement a PNR measurement (see Appendix \ref{appendix: threshold}). This method differs from conventional approaches by requiring only large pulse widths and post-processing analysis, without the need for additional detectors or delay lines (spatial and temporal multiplexing approaches) \cite{PhysRevLett.76.2464, PhysRevA.68.043814}. It enables PNR for arbitrarily high photon counts by increasing the pulse width, making it useful for characterization and proof-of-principle experiments. However, by increasing the pulse width and, consequently, the detection window, we increase the background noise and limit the communication rate. Therefore, in practice, a detector with PNR compatibility is preferable \cite{hao2024compact, Stasi:24}. We use the PNR measurement scheme to recover the message and compare it to the original, estimating the QBER by calculating the percentage of errors. By measuring the laser power at the emitter and accounting for channel and receiver losses, we can estimate the number of photons arriving at the detectors. This estimated value can be compared with the recorded clicks from the detectors for verification.

For the OOK implementation, we can opt to remove the polarization components (EOPM, PC, and PBS) or turning the laser off for one of the pulses. Only one detector connected to the time tagger is used. The setup is simple and compact, making it easy to transport or integrate into systems like a CubeSat for satellite quantum communication.

To implement both OOK-QKPC and PM-QKPC, a pulse width of 10 $\mu s$ and a rate of 50 kHz are chosen, accounting for the HVA limitations and enabling the use of PNR measurements. Messages are sent over 10 second intervals, repeated 10 times, with varying average photon numbers. Bob's average detector counts, $\eta |\alpha|^2$, is estimated based on the emitter's measured power and the total losses of the setup. This value is compared to the value measured by Bob, and in case of a mismatch, the experiment is repeated. The uncertainty in the QBER and average detector counts at the receiver is determined using the standard deviation. Photon noise is measured by turning off the laser, recording the photons detected per second, and calculating the average number per pulse.

\section{Results}
\label{sec:Results}

We implement both protocol encoding choices using the setup described in Section \ref{sec:experiment} and estimate the private capacity by applying equations derived in Section \ref{sec:concept} for different photon noise regimes.

\subsection{QBER}

\begin{figure}[h]
\centering
\begin{minipage}[t]{\linewidth}
a)
\centering
\includegraphics[width=0.85\textwidth]{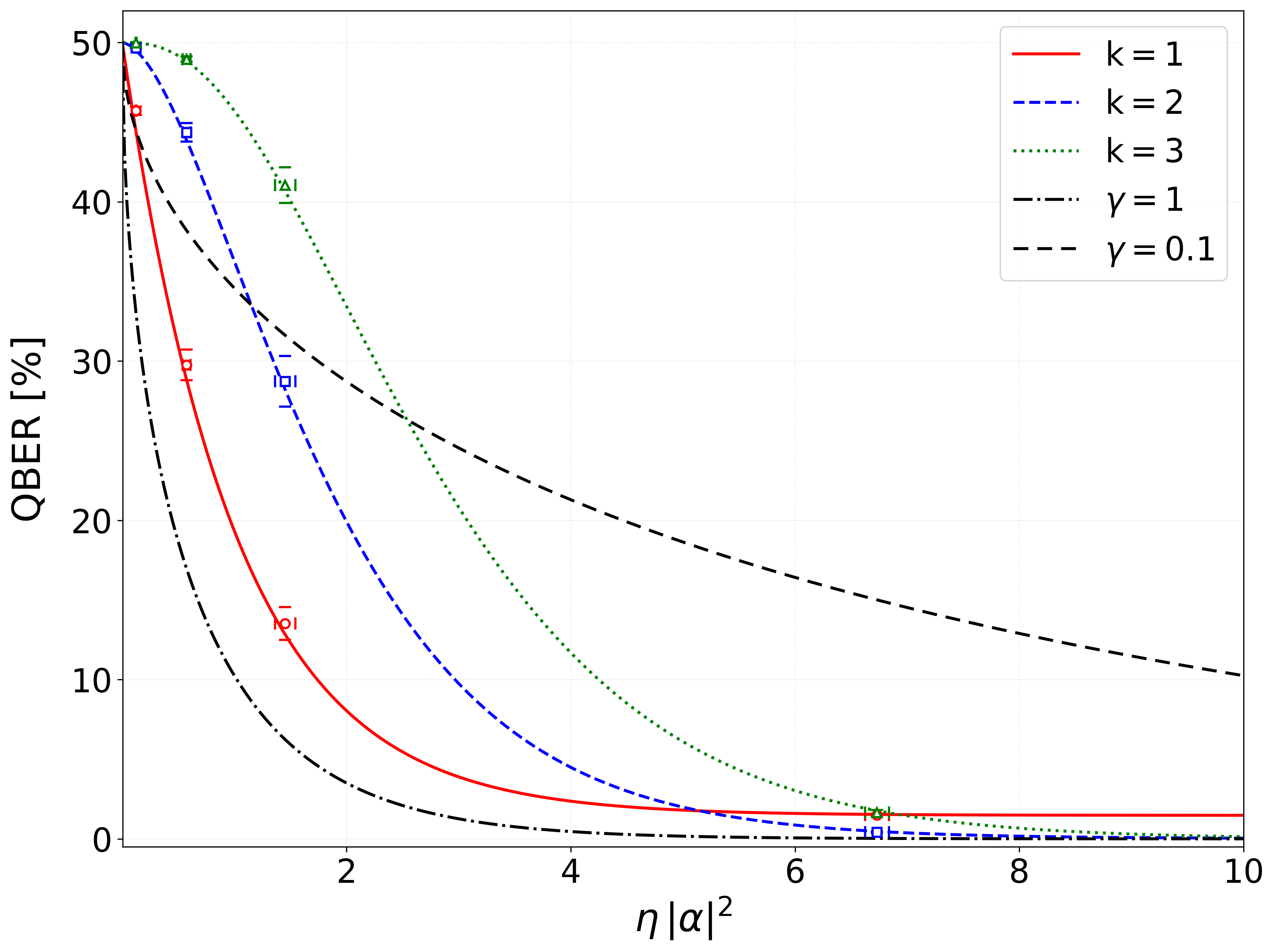} 

\end{minipage}

\begin{minipage}[t]{\linewidth}
b)
\centering
\includegraphics[width=0.8\textwidth]{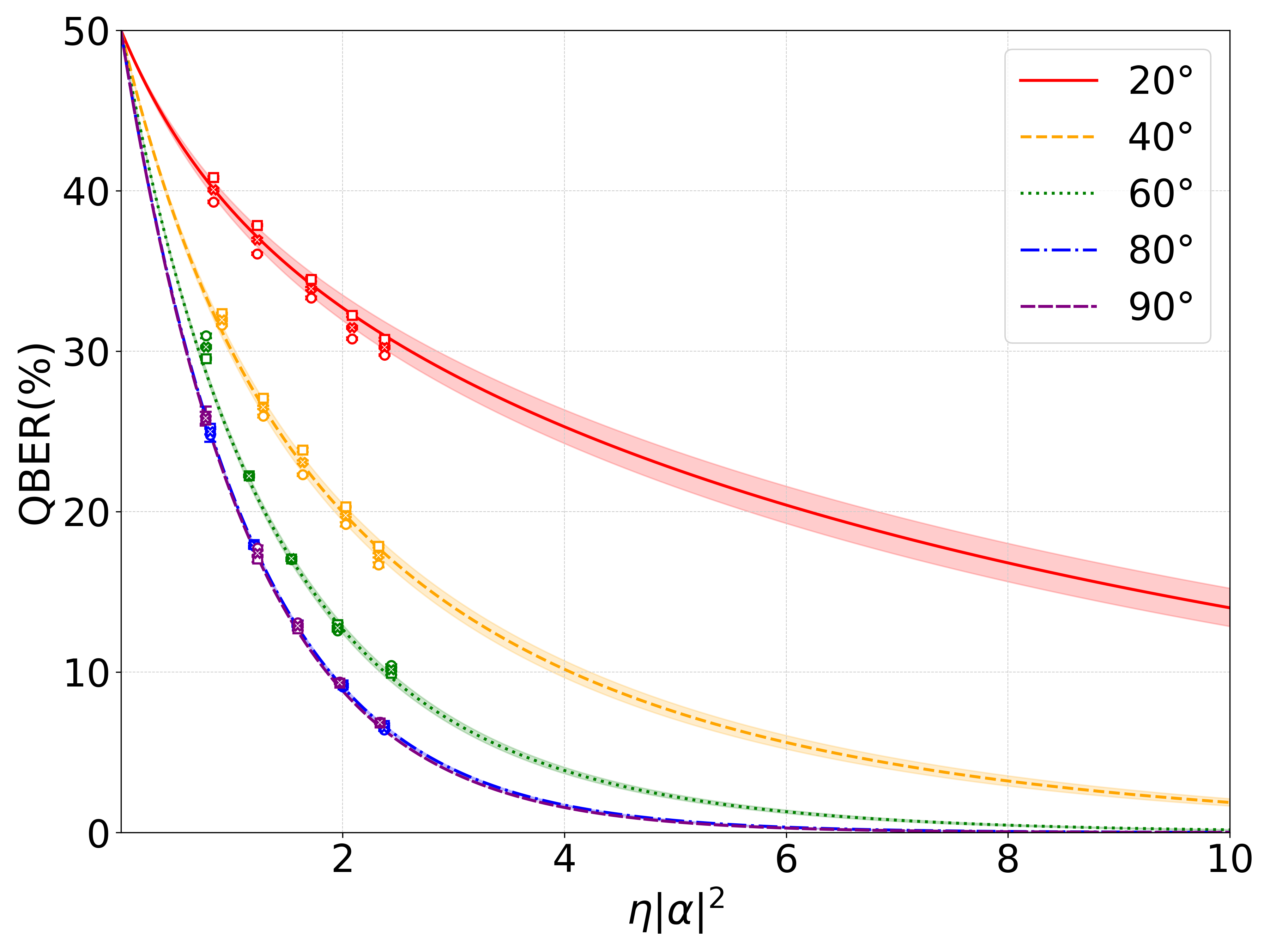}
\end{minipage}

\caption{QBER values as a function of the received number of photons with $\Delta = 0.03$ ($\approx 1500$ Hz). a) OOK-QKPC. In black we have Eve's QBER for different values of $\gamma$, and the colors show Bob's QBER for different discrimination strategies, different threshold choices, k. b) PM-QKPC. Different colors correspond to different polarization angles between the states. The markers correspond to different discrimination strategies (see Appendix \ref{appendix: discrimination}). The shaded area represents a one-degree error in the polarization preparation of the states.}
\label{fig:QBER}
\end{figure}

In Fig. \ref{fig:QBER}, the QBER as a function of the detected number of photons, $\eta |\alpha|^2$, for OOK and PM encoding can be seen. The experimental points shown were obtained as described in Section \ref{sec:experiment} and follow the expected curves for the different threshold choices. 

The QBER curves obtained from these non-ideal measurements are higher than the Helstrom bound if Eve receives the same number of photons $\gamma=1$. However, for $\gamma = 0.1$, where Eve intercepts only 10\% of the signal, her best achievable QBER remains above the practical measurements, allowing a positive private capacity and secure communication. This value is chosen as it offers a high private capacity while still being reasonable for practical implementation \cite{vazquez2021quantum}.

\subsection{Private Capacity}

To maximize the private capacity in Eq. \ref{eq: Private Capacity}, we optimize all tunable experimental variables: input distribution, the threshold value (for OOK), received photon number, polarization angle (for PM), and the relative difference in number of photons between both states (for PM) while holding the photon noise and $\gamma$ fixed. We found that the optimal input distribution is very close to uniform, and as this choice is easier for practical implementation, the results shown assume $q_0=\frac{1}{2}$, and that it is enough to consider two cases in the PM-QKPC: one state has no photons while the other has $|\alpha|^2$ (reduces to OOK) or both states share the same number of photons. For implementation purposes, we fix the minimum polarization angle to 2$\degree$.

\begin{figure}[ht]
\centering
\begin{minipage}[t]{\linewidth}
a)
\centering
\includegraphics[width=0.8\linewidth]{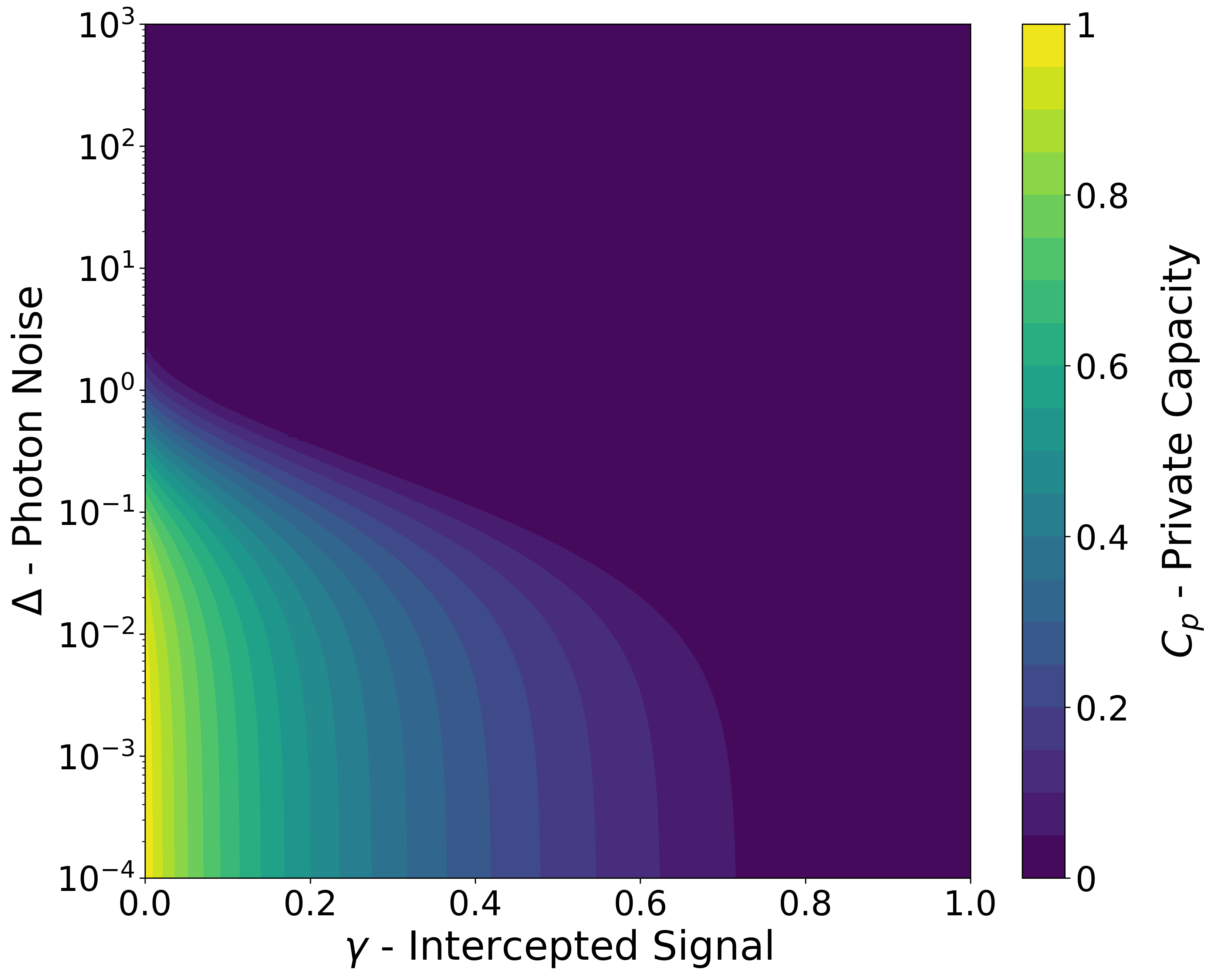}

\end{minipage}

\begin{minipage}[t]{\linewidth}
b)
\centering
\includegraphics[width=0.8\linewidth]{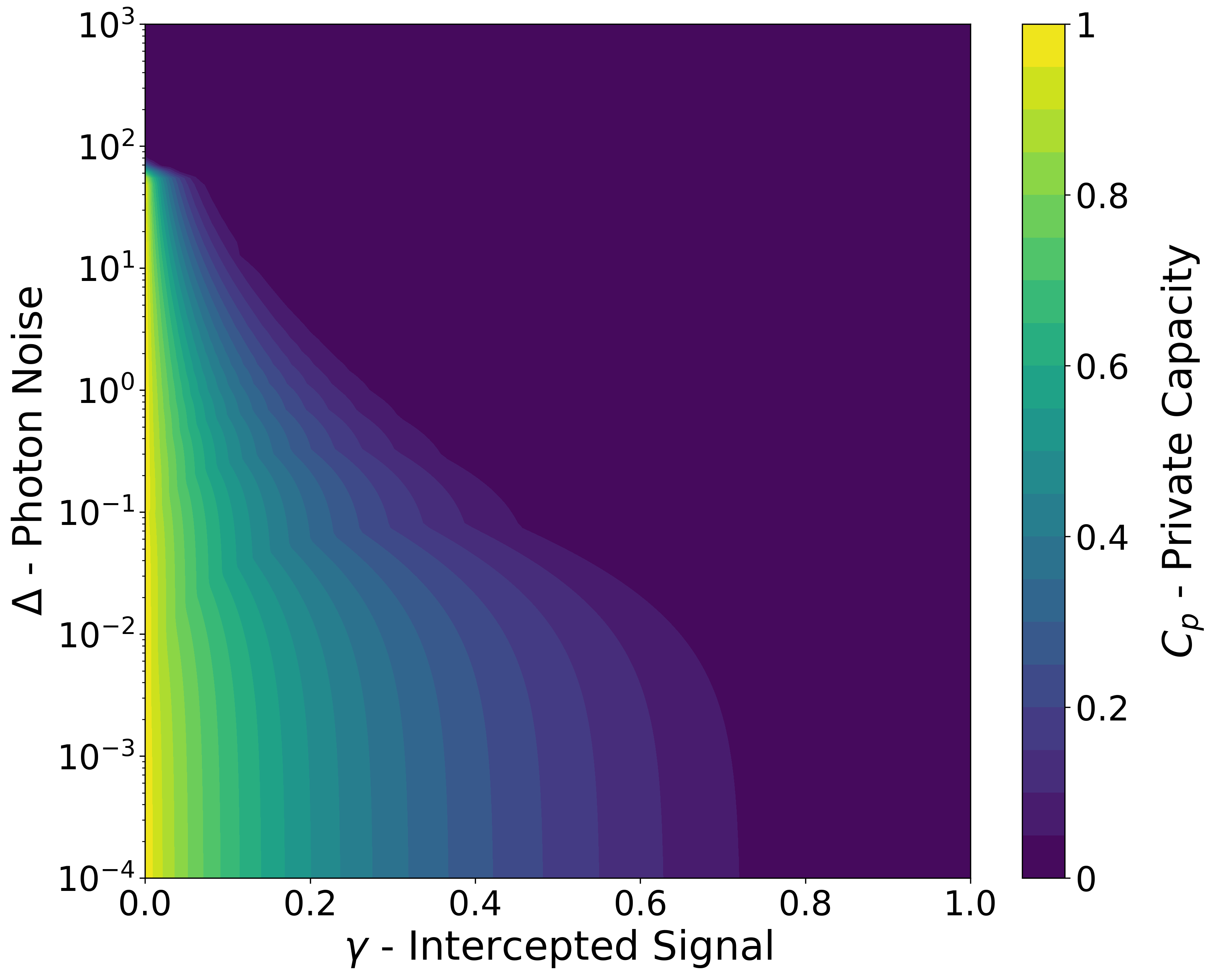}
\end{minipage}

\begin{minipage}[t]{\linewidth}
c)
\centering
\includegraphics[width=0.8\textwidth]{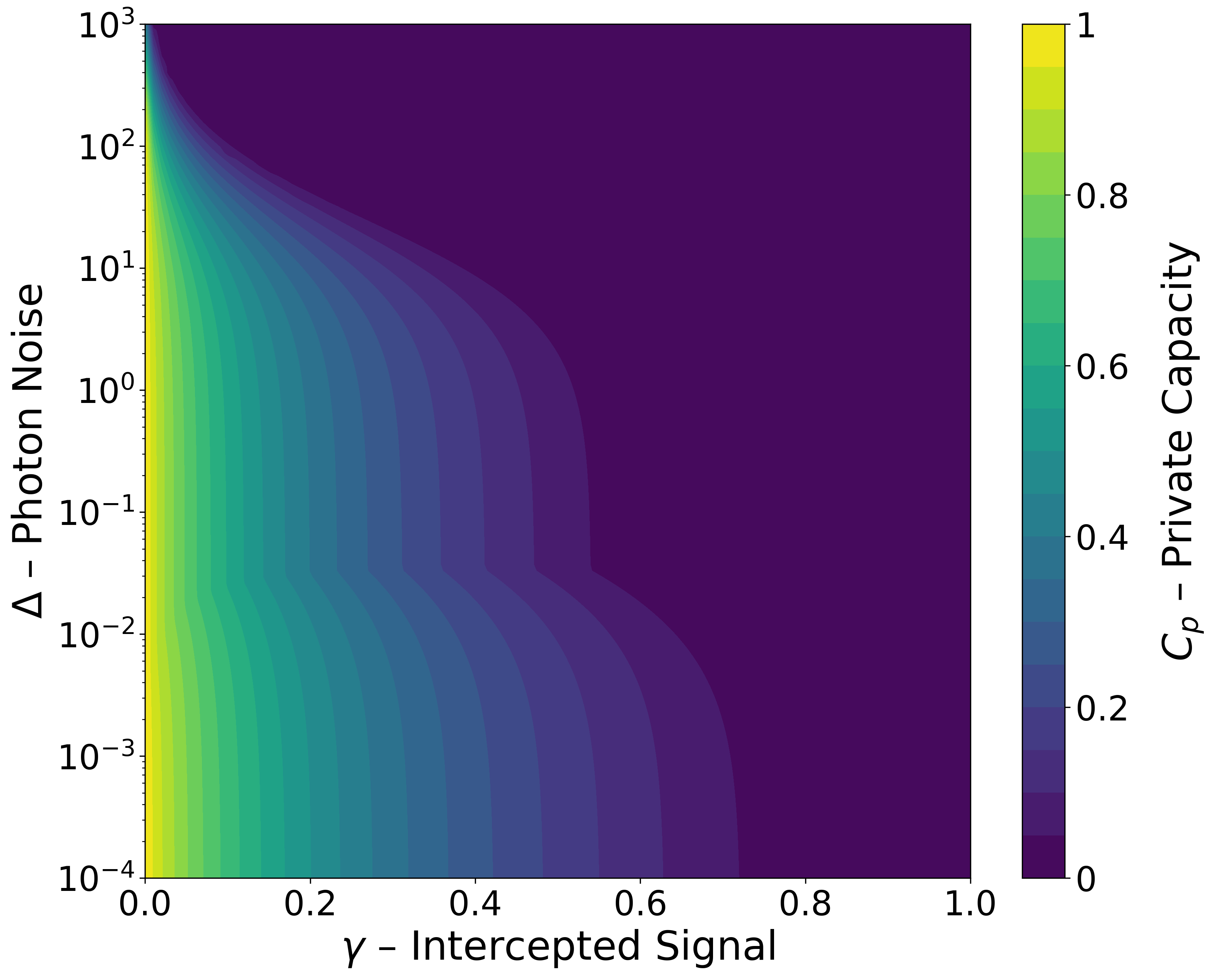}

\end{minipage}

\caption{Heat map of the maximum private capacity as a function of the photon noise, $\Delta$, and the intercepted signal by Eve, $\gamma$. a) OOK encoding, and a fixed threshold choice of $k=1$. b) OOK encoding, and a PNR detection scheme. c) PM encoding.}
\label{fig:heatmap_OOK}
\end{figure}

Fig. \ref{fig:heatmap_OOK} shows the maximum private capacity values for various levels of $\gamma$ and different photon noise regimes. For low photon noise levels, the results show no difference between using PNR measurements, non-PNR measurements, and PM-QKPC. In these conditions, for OOK-QKPC, setting $k=1$ for all $\gamma$ values is more advantageous, regardless of the measurement method and for PM-QKPC, the best strategy is turning off the laser. In this case, polarization does not improve state discrimination.

\begin{figure}[ht]
    \centering
    \includegraphics[width=0.40\textwidth]{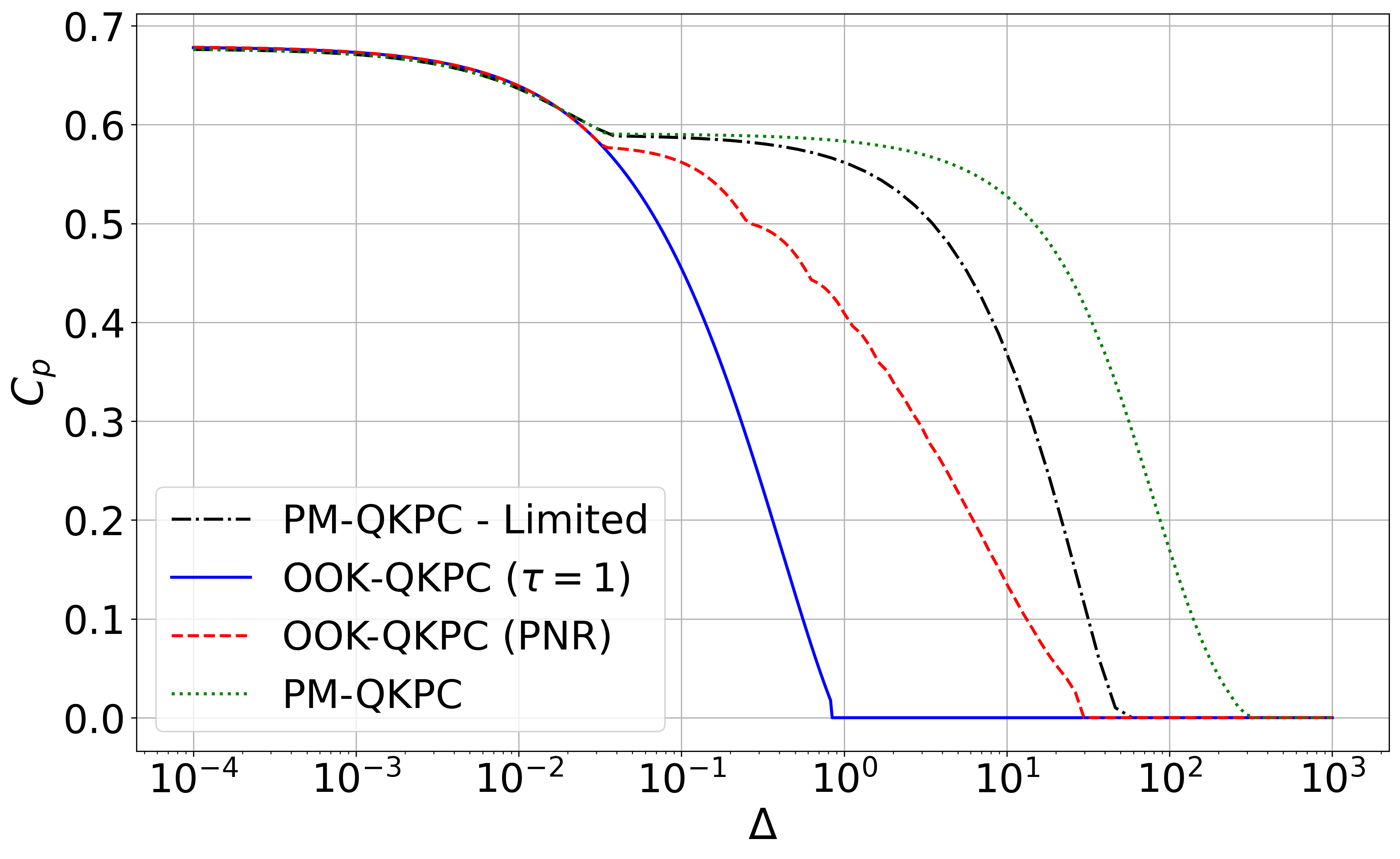} 
    \caption{Plot of the maximum private capacity as a function of the photon noise, $\Delta$, for $\gamma=0.1$ The different colors correspond to OOK-QKPC without PNR measurements (blue), OOK-QKPC with PNR measurements (red), PM-QKPC (green) and PM-QKPC with limited polarization angle and average number of photons (black).}
    \label{fig:2d_fixed_gamma}
\end{figure}

As photon noise increases, the private capacity for non-PNR measurements quickly drops to zero, whereas, for PNR measurements, the capacity decreases more gradually and remains nonzero for higher values, as seen in Fig \ref{fig:2d_fixed_gamma}. This can be explained by the fact that as the photon noise increases, the probability of detecting one or more photons when the laser is off (vacuum state) also increases, thereby introducing more errors into the discrimination process. To mitigate this, the threshold value must be increased to a level where the likelihood of unwanted photons introducing errors is minimized.

The PM-QKPC protocol provides two degrees of distinguishability: the difference in the average photon number between the states and their polarization. By turning off the laser for one state, we recover the results of OOK-QKPC, where discrimination relies entirely on the photon number difference. When the laser is on for both states, discrimination depends primarily on the polarization angle, making the results more robust against noise in the detector clicks.

As photon noise increases, PM-QKPC begins to outperform OOK-QKPC, even with PNR measurements. Using linearly polarized coherent states, we can control the overlap between the states, hence the intrinsic discrimination error. Therefore, Eve needs more light to distinguish between states, making it possible to communicate securely with a higher number of photons. This enables a near-maximum private capacity for fixed $\gamma$ at higher photon noise levels and maintains a positive private capacity, allowing secure communication, in much noisier conditions, as shown in Fig. \ref{fig:2d_fixed_gamma}. 

To achieve the maximum private capacity for high photon noise, a small polarization angle (below 10$\degree$) and a high average number of photons (above 100) is required, making it difficult to implement even for state-of-the-art photon counting detectors. This can be solved by limiting the average number of photons and the polarization angle to a practical value (20 and 10$\degree$, for example) and finding the maximum private capacity for that case as shown in black in Fig. \ref{fig:2d_fixed_gamma}. While the private capacity decreases, it still allows for higher values than OOK-QKPC for high photon noise regimes.

\section{Comparison with QKD}
\label{sec:Comparison}

We would like to point out that a direct comparison between QKPC and QKD is difficult because of the very different nature of the cryptographic scenarios and assumptions. QKD is used to generate secret keys, which are stored and later consumed by classical encryption schemes, while QKPC enables the direct transmission of messages without the need for key management or key storage. As a result, secret key rates and secure message transmission rates should not be interpreted as equivalent performance metrics.

The two approaches also rely on distinct assumptions regarding an eavesdropper's capabilities. In QKD, Eve is assumed to have unrestricted access to the quantum channel, with security guaranteed through the detection of disturbances induced by measurement. In contrast, QKPC is based on a wiretap channel model, in which security is ensured under the assumption that Eve's channel is bounded relative to that of Bob. 

Nevertheless, a comparison of achievable rates remains interesting when performed under the same physical conditions, particularly in regimes where QKD performance is known to be limited (daylight). We focus on free-space line-of-sight implementations, as the same physical channel and hardware could, in principle, be used for both QKD and QKPC.

The bound on the parameter $\gamma$ is crucial to any implementation of QKPC. While it is an assumption inherent to the wiretap channel model, it is important to use physical constraints to estimate it. In free-space scenarios, geometric considerations such as beam divergence, receiver aperture size, and exclusion zones around the receiver naturally limit the fraction of optical power that Eve can collect. Additionally, practical deployments may employ channel monitoring strategies, such as spatial filtering, beam dumps, or surveillance of the propagation path, to prevent or detect disturbances. While these measures do not provide the intrinsic intrusion detection offered by QKD, they allow $\gamma$ to be conservatively bounded in realistic settings without introducing additional assumptions beyond the standard wiretap model.

For the comparison presented in this section, we adopt the same representative value of $\gamma=0.1$ used in previous work \cite{vazquez2021quantum}, where explicit physical models for space-based line-of-sight scenarios were introduced and discussed.

In \cite{avesani2021full}, a 145 m free space link is established in the urban area of Padua, Italy. QKD is used to generate keys at an average rate of 30 kbps under daylight conditions. The source rate is 50 MHz with 500 ps of full-width-at-half-maximum pulses.  Commercially available superconductive nanowire single-photon detectors (SNSPDs) are used to detect weak quantum states of light. Using different filtering techniques, (see Appendix \ref{appendix: background}), the authors of that article achieve a photon noise of 240 Hz or $1
4.8\times10^{-6}$ per pulse.


In the same experimental conditions, same detectors, channel parameters, photon noise, and assuming $\gamma = 0.1$, the secret capacity is around 0.66 and enables communication at a secure rate of 33 Mbps. Assuming an even larger background noise, with OOK-QKPC (using the same single photon detectors as \cite{avesani2021full}), we can still achieve a 1 Mbps communication rate up to $0.8$ background photons per pulse. 

If we consider the use of detectors with PNR capabilities like the ones in \cite{hao2024compact, Stasi:24} and keep the source parameters, we can communicate faster than 1 Mbps even with an increase in the photon noise up to 30 photons per pulse. 

\section{Discussion}
\label{sec:conclusion}

In this work, we have introduced a polarization-based QKPC protocol, which we call PM-QKPC, along with a photon number resolving measurement scheme based on threshold detectors. The use of the measurement scheme considerably enhances the original protocol, OOK-QKPC, in terms of resistance to background noise. The use of polarization encoding further improves the resilience, enabling robust daylight operation of QKPC. A thorough comparison with QKD shows an advantage of a factor four in terms of the tolerated background photon number per pulse.

QKPC is promising for free-space quantum communication applications, including space-based quantum communication. It enables quantum secure high rate data transfer with moderate experimental requirements. Future research directions include studying the scaling of QKPC performance with distance, under turbulent conditions, and creating portable miniaturized systems.


\begin{acknowledgments}
This work is funded by national funds through FCT – Fundação para a Ciência e a Tecnologia, I.P., and, when eligible, co-funded by EU funds under project/support UID/50008/2025 – Instituto de Telecomunicações, with DOI identifier https://doi.org/10.54499/UID/50008/2025. The authors thank the support from the European Commission (EC) through project PTQCI (DIGITAL-2021-QCI-01). This work was supported by the Galician Regional Government (consolidation of Research Units: AtlantTIC), MICIN with funding from the European Union NextGenerationEU (PRTR-C17.I1) and the
Galician Regional Government with own funding
through the ”Planes Complementarios de I+D+I con
las Comunidades Autónomas” in Quantum Communication
and the European Union’s Horizon Europe
Framework Programme under the project ”Quantum
Security Networks Partnership” (QSNP, grant agreement
No101114043). P.N.M. acknowledges the support of FCT through scholarship 2024.01717.BD.
\end{acknowledgments}

\bibliography{apssamp}

@article{endo2016free,
  title={Free-space optical channel estimation for physical layer security},
  author={Endo, Hiroyuki and Fujiwara, Mikio and Kitamura, Mitsuo and Ito, Toshiyuki and Toyoshima, Morio and Takayama, Yoshihisa and Takenaka, Hideki and Shimizu, Ryosuke and Laurenti, Nicola and Vallone, Giuseppe and others},
  journal={Optics express},
  volume={24},
  number={8},
  pages={8940--8955},
  year={2016},
  publisher={Optical Society of America}
}

@article{poor2017wireless,
  title={Wireless physical layer security},
  author={Poor, H Vincent and Schaefer, Rafael F},
  journal={Proceedings of the National Academy of Sciences},
  volume={114},
  number={1},
  pages={19--26},
  year={2017},
  publisher={National Academy of Sciences}
}

@inproceedings{Stasi:24,
author = {Lorenzo Stasi and Towsif Taher and Giovanni V. Resta and Hugo Zbinden and Rob Thew and F\'{e}lix Bussi\`{e}res},
booktitle = {Quantum 2.0 Conference and Exhibition},
journal = {Quantum 2.0 Conference and Exhibition},
pages = {QM2A.2},
publisher = {Optica Publishing Group},
title = {High-efficiency photon-number resolution and 250 Mcps detection rate with a 28-pixel superconducting nanowire single-photon detector},
year = {2024},
url = {https://opg.optica.org/abstract.cfm?URI=QUANTUM-2024-QM2A.2},
doi = {10.1364/QUANTUM.2024.QM2A.2},
}

@article{zheng2025free,
  title={Free-space continuous-variable quantum key distribution under high background noise},
  author={Zheng, Xue-Tao and Zhang, Qi-Fa and Ling, Jie and Guo, Guang-Can and Han, Zheng-Fu},
  journal={npj Quantum Information},
  volume={11},
  number={1},
  pages={52},
  year={2025},
  publisher={Nature Publishing Group UK London}
}

@article{cai2024free,
  title={Free-space quantum key distribution during daylight and at night},
  author={Cai, Wen-Qi and Li, Yang and Li, Bo and Ren, Ji-Gang and Liao, Sheng-Kai and Cao, Yuan and Zhang, Liang and Yang, Meng and Wu, Jin-Cai and Li, Yu-Huai and others},
  journal={Optica},
  volume={11},
  number={5},
  pages={647--652},
  year={2024},
  publisher={Optica Publishing Group}
}

@article{ko2018experimental,
  title={Experimental filtering effect on the daylight operation of a free-space quantum key distribution},
  author={Ko, Heasin and Kim, Kap-Joong and Choe, Joong-Seon and Choi, Byung-Seok and Kim, Jong-Hoi and Baek, Yongsoon and Youn, Chun Ju},
  journal={Scientific reports},
  volume={8},
  number={1},
  pages={15315},
  year={2018},
  publisher={Nature Publishing Group UK London}
}

@article{abasifard2024ideal,
  title={The ideal wavelength for daylight free-space quantum key distribution},
  author={Abasifard, Mostafa and Cholsuk, Chanaprom and Pousa, Roberto G and Kumar, Anand and Zand, Ashkan and Riel, Thomas and Oi, Daniel KL and Vogl, Tobias},
  journal={APL Quantum},
  volume={1},
  number={1},
  year={2024},
  publisher={AIP Publishing}
}

@article{liao2017long,
  title={Long-distance free-space quantum key distribution in daylight towards inter-satellite communication},
  author={Liao, Sheng-Kai and Yong, Hai-Lin and Liu, Chang and Shentu, Guo-Liang and Li, Dong-Dong and Lin, Jin and Dai, Hui and Zhao, Shuang-Qiang and Li, Bo and Guan, Jian-Yu and others},
  journal={Nature Photonics},
  volume={11},
  number={8},
  pages={509--513},
  year={2017},
  publisher={Nature Publishing Group UK London}
}

@article{avesani2021full,
  title={Full daylight quantum-key-distribution at 1550 nm enabled by integrated silicon photonics},
  author={Avesani, M and Calderaro, L and Schiavon, M and Stanco, A and Agnesi, C and Santamato, A and Zahidy, M and Scriminich, A and Foletto, G and Contestabile, G and others},
  journal={npj Quantum Information},
  volume={7},
  number={1},
  pages={93},
  year={2021},
  publisher={Nature Publishing Group UK London}
}

@article{er2005background,
  title={Background noise of satellite-to-ground quantum key distribution},
  author={Er-Long, Miao and Zheng-fu, Han and Shun-sheng, Gong and Tao, Zhang and Da-Sheng, Diao and Guang-Can, Guo},
  journal={New Journal of Physics},
  volume={7},
  number={1},
  pages={215},
  year={2005},
  publisher={IOP Publishing}
}

@article{vazquez2024quantum,
  title={Quantum Keyless Private Communication with Decoy States for Space Channels},
  author={V{\'a}zquez-Castro, {\'A}ngeles and Winter, Andreas and Zbinden, Hugo},
  journal={IEEE Transactions on Information Forensics and Security},
  year={2024},
  publisher={IEEE}
}

@PREAMBLE{
 "\providecommand{\noopsort}[1]{}" 
 # "\providecommand{\singleletter}[1]{#1}%" 
}

@article{hao2024compact,
  title={A compact multi-pixel superconducting nanowire single-photon detector array supporting gigabit space-to-ground communications},
  author={Hao, Hao and Zhao, Qing-Yuan and Huang, Yang-Hui and Deng, Jie and Yang, Fan and Ru, Sai-Ying and Liu, Zhen and Wan, Chao and Liu, Hao and Li, Zhi-Jian and others},
  journal={Light: Science \& Applications},
  volume={13},
  number={1},
  pages={25},
  year={2024},
  publisher={Nature Publishing Group UK London}
}

@article{vazquez2021quantum,
  title={Quantum keyless private communication versus quantum key distribution for space links},
  author={V{\'a}zquez-Castro, A and Rusca, Davide and Zbinden, Hugo},
  journal={Physical Review Applied},
  volume={16},
  number={1},
  pages={014006},
  year={2021},
  publisher={APS}
}

@article{wyner1975wire,
  title={The wire-tap channel},
  author={Wyner, Aaron D},
  journal={Bell system technical journal},
  volume={54},
  number={8},
  pages={1355--1387},
  year={1975},
  publisher={Wiley Online Library}
}

@article{wyner1988capacity2,
  title={Capacity and error exponent for the direct detection photon channel. II},
  author={Wyner, Aaron D},
  journal={IEEE Transactions on Information Theory},
  volume={34},
  number={6},
  pages={1462--1471},
  year={1988},
  publisher={IEEE}
}

@article{pan2023free,
  title={Free-space quantum secure direct communication: Basics, progress, and outlook},
  author={Pan, Dong and Song, Xiao-Tian and Long, Gui-Lu},
  journal={Advanced Devices \& Instrumentation},
  volume={4},
  pages={0004},
  year={2023},
  publisher={AAAS}
}

@article{PhysRevA.68.043814,
  title = {Photon-number resolution using time-multiplexed single-photon detectors},
  author = {Fitch, M. J. and Jacobs, B. C. and Pittman, T. B. and Franson, J. D.},
  journal = {Phys. Rev. A},
  volume = {68},
  issue = {4},
  pages = {043814},
  numpages = {6},
  year = {2003},
  month = {Oct},
  publisher = {American Physical Society},
  doi = {10.1103/PhysRevA.68.043814},
  url = {https://link.aps.org/doi/10.1103/PhysRevA.68.043814}
}

@article{PhysRevLett.76.2464,
  title = {Photon Chopping: New Way to Measure the Quantum State of Light},
  author = {Paul, H. and T\"orm\"a, P. and Kiss, T. and Jex, I.},
  journal = {Phys. Rev. Lett.},
  volume = {76},
  issue = {14},
  pages = {2464--2467},
  numpages = {0},
  year = {1996},
  month = {Apr},
  publisher = {American Physical Society},
  doi = {10.1103/PhysRevLett.76.2464},
  url = {https://link.aps.org/doi/10.1103/PhysRevLett.76.2464}
}

@article{mendes2024optical,
  title={Optical payload design for downlink quantum key distribution and keyless communication using CubeSats},
  author={Mendes, Pedro Neto and Teixeira, Gon{\c{c}}alo Lobato and Pinho, David and Rocha, Rui and Andr{\'e}, Paulo and Niehus, Manfred and Faleiro, Ricardo and Rusca, Davide and Cruzeiro, Emmanuel Zambrini},
  journal={EPJ Quantum Technology},
  volume={11},
  number={1},
  pages={48},
  year={2024},
  publisher={Springer Berlin Heidelberg}
}

@article{berra2023synchronization,
  title={Synchronization of quantum communications over an optical classical communications channel},
  author={Berra, Federico and Agnesi, Costantino and Stanco, Andrea and Avesani, Marco and Kuklewski, Michal and Matter, Daniel and Vallone, Giuseppe and Villoresi, Paolo},
  journal={Applied Optics},
  volume={62},
  number={30},
  pages={7994--7999},
  year={2023},
  publisher={Optica Publishing Group}
}

@article{calderaro2020fast,
  title={Fast and simple qubit-based synchronization for quantum key distribution},
  author={Calderaro, Luca and Stanco, Andrea and Agnesi, Costantino and Avesani, Marco and Dequal, Daniele and Villoresi, Paolo and Vallone, Giuseppe},
  journal={Physical Review Applied},
  volume={13},
  number={5},
  pages={054041},
  year={2020},
  publisher={APS}
}

@article{boaron2018secure,
  title={Secure quantum key distribution over 421 km of optical fiber},
  author={Boaron, Alberto and Boso, Gianluca and Rusca, Davide and Vulliez, C{\'e}dric and Autebert, Claire and Caloz, Misael and Perrenoud, Matthieu and Gras, Ga{\"e}tan and Bussi{\`e}res, F{\'e}lix and Li, Ming-Jun and others},
  journal={Physical review letters},
  volume={121},
  number={19},
  pages={190502},
  year={2018},
  publisher={APS}
}

\clearpage

\appendix
\section*{Appendices}

\section{Private capacity estimation for QKPC}
\label{appendix: discrimination}

\begin{figure}[ht]
    \centering
    \includegraphics[width=0.35\textwidth]{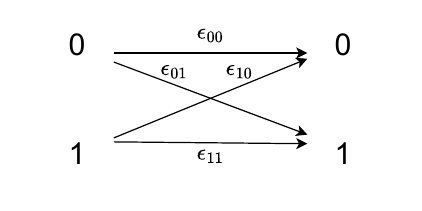}
    \caption{Diagram illustrating the notation for a binary communication channel.}
    \label{fig:binary channel}
\end{figure}

For both the OOK and PM implementation of the QKPC protocol, an asymmetric binary communication channel is assumed, Fig. \ref{fig:binary channel}. 

\subsection{OOK-QKPC}

We begin by describing OOK-QKPC using photon number resolving (PNR) measurements. We consider a binary channel with input probabilities $q_0$ and $q_1 = 1 - q_0$. Depending on the input bit $x\in \{0,1\}$, the prepared state is either the vacuum $|\psi_{x=0} \rangle \equiv |\psi_0 \rangle = |0\rangle$ or a coherent state $|\psi_{x=1} \rangle \equiv |\psi_1 \rangle = |\alpha\rangle$, prepared as in Eq. \ref{Eq: OOK states_app}, a vacuum state and a general weak coherent state.

\begin{equation}
\begin{aligned}
    |\psi_0 \rangle &= |0\rangle, \\
    |\psi_1 \rangle &= |\alpha \rangle.
\end{aligned}
\label{Eq: OOK states_app}
\end{equation}

The receiver uses PNR detectors to measure the incoming states. For each state sent, the detector counts the number of arriving photons. To discriminate the states, we use a simple strategy: if the detector registers a number of photons equal or greater than a given threshold, 
k, we assume the state sent was $|\psi_1 \rangle$ (as this state has a higher average photon number than $|\psi_0 \rangle$), and the bit recorded is 1. If the detector measures a number of photons lower than k, the bit recorded is 0. For k$=1$, no PNR capabilities are necessary, threshold single-photon detectors are sufficient.

Using this strategy, we can characterize the binary communication channel. We assume there are $\Delta$ detector clicks per pulse on average from background photons or/and dark counts, photon noise, and these clicks in the detector follow a Poisson distribution and are independent of the emitter signal. The probability of sending bit 0 and recording bit 0 is given by the probability of sending state $|\psi_0 \rangle$ and measuring fewer than k photons. This probability is given by:

\begin{equation}
    \epsilon_{00} = \sum_{i=0}^{k - 1} e^{-\Delta} \frac{\Delta^i}{i!}.
\end{equation}

The probability of sending bit 1 and recording bit 0 is given by the probability of sending state $|\psi_1 \rangle$ and measuring fewer than k photons. This probability is given by:

\begin{equation}
    \epsilon_{10} = \sum_{i=0}^{k - 1} e^{-(\eta|\alpha|^2 +\Delta)} \frac{(\eta|\alpha|^2 +\Delta)^i}{i!}.
\end{equation}

The binary communication channel is then fully defined as $\epsilon_{01} = 1 - \epsilon_{00}$ and $\epsilon_{11} = 1 - \epsilon_{10}$. 

The probability of a discrimination error for a chosen k is:

\begin{equation}
    \text{QBER} = \frac{\epsilon_{01} + \epsilon_{10}}{2},
    \label{eq: QBER_app}
\end{equation}

where we consider a uniform input distribution (i.e. $q_0$ equals $q_1$) as we found it to be very close to the optimal value.
The mutual information for Bob, $I_B$, is given by:

\begin{equation}
    I_B = h_b\left(\frac{\epsilon_{00} + \epsilon_{10}}{2}\right) - \frac{h_b(\epsilon_{00}) + h_b(\epsilon_{10})}{2},
    \label{eq: Mutual_information_Bob_app}
\end{equation}

where $h_b$ is the binary entropy function.
Eve is assumed to have a binary symmetric channel with error probability $\epsilon_{\gamma}$, given by the Helstrom bound. We assume Eve only intercepted a fraction, $\gamma$, of the signal Bob received. For the states assumed in Eq. \ref{Eq: OOK states},

\begin{equation}
    \epsilon_{\gamma} = \frac{1}{2}\left(1-\sqrt{1-|\langle\psi_0|\psi_1\rangle|^2}\right),
    \label{eq: Qber_helstrom_app}
\end{equation}

with

\begin{equation}
    \langle\psi_0|\psi_1\rangle = e^{-\gamma\eta\frac{|\alpha|^2}{2}}.
\end{equation}

For orthogonal states, the overlap is zero, allowing Eve to perfectly discriminate between them without error, rendering the communication entirely insecure.

Eve's mutual information, $I_E$, is then:

\begin{equation}
    I_E = 1 - h_b(\epsilon_{\gamma}),
    \label{eq: Mutual_information_Eve_app}
\end{equation}

and the private capacity is:

\begin{equation}
    C_p(\gamma) = \text{max}_{|\alpha|^2,\tau} \left[I_B - I_E\right]
    \label{eq: Private Capacity_app}
\end{equation}

If counting the number of single photons arriving at the detector is not possible, only a threshold detector is available, the only threshold choice possible will be k$ = 1$.

\subsection{PM-QKPC}

For the PM implementation, two general weak coherent states with polarization in the XZ plane are prepared as in Eq. \ref{Eq: Polarization states}, state $|\psi_0 \rangle$ ($|\psi_1 \rangle$) for the input bit 0 (1).

\begin{equation}
\begin{aligned}
    |\psi_0 \rangle &= |\alpha\rangle_H \otimes |0\rangle_V, \\
    |\psi_1 \rangle &= |\cos\theta\beta\rangle_H \otimes |\sin\theta\beta\rangle_V.
\end{aligned}
\label{Eq: Polarization states}
\end{equation}

To maximize the discrimination probability, the same rotation explained in the main text is considered. The state of each output is measured by a single photon detector and a photon noise per pulse at each detector is considered, $\Delta$, as in the OOK encoding. For state $|\psi_0 \rangle$, the average number of photons that arrive at each detector from the emitter is: 

\begin{equation}
\begin{aligned}
    \text{Detector 0: } \cos^2 \left( \delta \right) \eta| \alpha |^2 = \eta| \alpha_{H} |^2 \\
    \text{Detector 1: } \sin^2 \left( \delta \right) \eta| \alpha |^2 = \eta| \alpha_{V} |^2,
\end{aligned}
\end{equation}

and for state $|\psi_1 \rangle$:

\begin{equation}
\begin{aligned}
   \text{Detector 0: } \cos^2 \left( \theta + \delta \right) \eta| \beta |^2 = \eta| \beta_{H} |^2 \\
   \text{Detector 1: } \sin^2 \left( \theta + \delta \right) \eta| \beta |^2 = \eta| \beta_{V} |^2.
\end{aligned}
\end{equation}

We define a ratio between the number of photons of both states, $\kappa =|\beta |^2/|\alpha |^2$.

As $\cos^2 \left( \delta \right) = \sin^2 \left( \theta + \delta \right)$ and $\sin^2 \left( \delta \right) = \cos^2 \left( \theta + \delta \right)$, $\kappa | \alpha_{H} |^2 = | \beta_{V} |^2$ and $\kappa | \alpha_{V} |^2 = | \beta_{H} |^2$.

Using the same strategy as explained in the main text, the probability of a difference of $m$ clicks between both detectors is given by the convolution of each detector distribution conditioned on the input bit:

\begin{equation}
\begin{split}
P(m|0) &= e^{-\left(\eta|\alpha |^2 + 2\Delta\right)}
\sum_{l=0}^{\infty}
\frac{(\eta|\alpha_{H} |^2 + \Delta)^{l+m}}
{(l+m)! \, l!} \\
&\quad \times (\eta|\alpha_{V} |^2 + \Delta)^l .
\end{split}
\end{equation}

\begin{equation}
\begin{split}
P(m|1) &= e^{-\left(\kappa \eta|\alpha |^2 + 2\Delta\right)}
\sum_{l=0}^{\infty}
\frac{(\kappa \eta|\alpha_{V} |^2 + \Delta)^{l+m}}
{(l+m)! \, l!} \\
&\quad \times (\kappa \eta|\alpha_{H} |^2 + \Delta)^l .
\end{split}
\end{equation}

The rest of the analysis is done in the same way. When maximizing the private capacity over the input parameters, we find either $\kappa \approx 0$ (recovering the results from the OOK QKPC) or $\kappa \approx 1$. In addition, the optimal input distribution is very close to uniform.

It is important to also define a strategy for when both detectors click the same number of times. We consider three discrimination strategies: always choose bit 0, always choose bit 1, or randomly select between the two.

The probability of observing an equal number of photons in each detector is denoted by $P(m=0|0)$ for bit 0 and $P(m=0|1)$ for bit 1. Consequently, the error probabilities associated with each decision strategy are as follows:
\begin{itemize}
    \item If we always choose bit 0 when the number of photons is equal, the error probability is $P(m=0|1)$, the probability of incorrectly choosing bit 0 when bit 1 was sent.
    \item If we always choose bit 1 when the number of photons is equal, the error probability is $P(m=0|0)$, the probability of incorrectly choosing bit 1 when bit 0 was sent.
    \item If we randomly choose the bit following the input distribution (assuming uniform distribution), the error probability is $\frac{1}{2}$$P(m=0|1)$ + $\frac{1}{2}$$P(m=0|0)$.
\end{itemize}

The probabilities $P(m=0|0)$ and $P(m=0|1)$ are given by the following expressions:

\begin{equation}
    P(m=0|0) = e^{-(\eta|\alpha |^2)} \sum_{l=0}^{\infty} \frac{(\eta|\alpha_{0} |^2  \eta|\alpha_{1} |^2)^{l}}{(l!)^2},
\end{equation} 

\begin{equation}
    P(m=0|1) = e^{-\left(\kappa \eta|\alpha |^2\right)} \sum_{l=0}^{\infty} \frac{(\kappa^2 \eta|\alpha_{0} |^2 |\alpha_{1} |^2)^l}{(l!)^2}.
\end{equation} 


For $\kappa \in [0,1]$, we have $P(m=0|1) \geq P(m=0|0)$. This is expected as the state with a lower number of photons will have a higher probability of making the detectors click the same number of times. This means that the optimal discrimination strategy is to assume that all events with the same number of clicks on both detectors correspond to the case $x=1$.

\section{Validity of approximation of time-multiplexed threshold detectors as PNR measurements}
\label{appendix: threshold}

For a general coherent state $|\alpha \rangle$, with $|\alpha|^2$ average number of photons, the probability of detecting n photons is given by the Poisson distribution: 

\begin{equation}
    P(n) = \left| \langle n | \alpha \rangle \right|^2 = e^{-|\alpha|^2} \frac{(|\alpha|^2)^n}{n!}.
\end{equation}

Threshold single-photon detectors are limited to measuring the presence or absence of photons within a given measurement interval, which is constrained by the detector's dead time. The dead time of a detector refers to the period immediately following a detection event during which the detector is unable to register another photon. Consequently, these detectors cannot capture all the information contained in a coherent state in a single measurement

A threshold detector measurement is described by the following Positive Operator-Valued Measure (POVM) consisting of two elements:

\begin{align}
\hat{\Pi}^0 & = |0\rangle\langle0|\\
\hat{\Pi}^1 & = \sum_{j=1} |j\rangle\langle j| = \mathbf{I} - |0\rangle\langle0|
\end{align}

\begin{figure}[ht]
    \centering
    \includegraphics[width=0.35\textwidth]{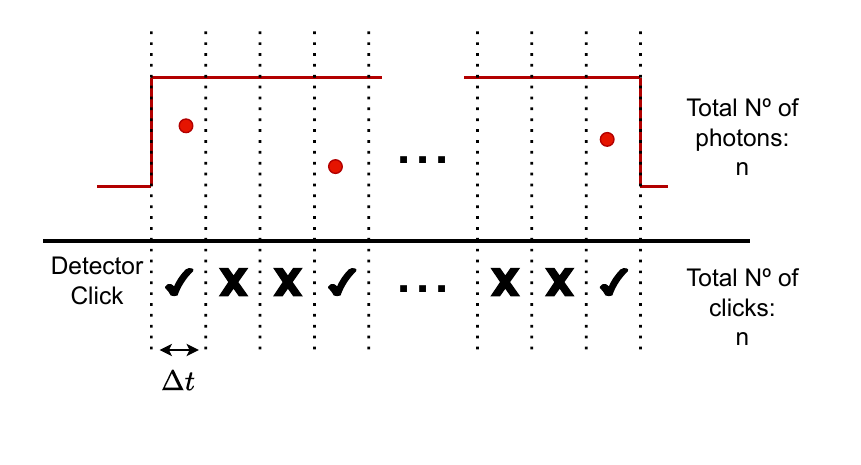} 
    \caption{Illustration of the time multiplexing technique to count the number of photons using threshold detectors. $\Delta t$ is the dead time of the detector.}
    \label{fig:PNR_ilustration}
\end{figure}

Supposing we have a very long pulse of light and a small number of photons relative to the dead time of the detector, illustrated in Fig. \ref{fig:PNR_ilustration}, the detector will register a succession of clicks (detections) and no-clicks (no detections) over time. This series of measurements with a threshold detector can be represented by the global measurement,

\begin{equation}
\hat{\Pi}^{\mathbf{d}}_i = \bigotimes_{i=1}^{N}  \hat{\Pi}^{d_i}_i,
\end{equation}

where $N$ is the number of measurement intervals, and the elements $d_i$ of the vector $\mathbf{d} $ represent the measurement outcome for each interval $i = 1, \dots, N $.

Assuming, without loss of generality, that we have a set $\mathbb{Z}_0 $ of measurements where $d_i = 0 $ and a set $ \mathbb{Z}_1 $ where $ d_i = 1 $, then

\begin{equation}
\hat{\Pi}^{\mathbf{d}}_i = \bigotimes_{l \in \mathbb{Z}_0}  \hat{\Pi}^{0}_l \bigotimes_{k \in \mathbb{Z}_1}  \hat{\Pi}^{1}_k.
\end{equation}

As there are only two outcomes and the measurements are independent, the probability of obtaining a total number of $ k $ photons follows a binomial distribution with probability $p_i$ of detecting at least one photon in a given interval,

\begin{equation}
p = 1 - \langle \alpha_i |\hat{\Pi}^{0}| \alpha_i \rangle = 1 - \langle \alpha_i | 0  \rangle \langle 0 | \alpha_i \rangle = 1- e^{-\frac{|\alpha|^2}{N}},
\end{equation}

and 

\begin{equation}
P(X = k) = \binom{N}{k} p^k (1-p)^{N-k}.
\end{equation}

In probability theory, the law of rare events, or Poisson limit theorem, states that the Poisson distribution may be used as an approximation to the binomial distribution, for small number of events in relation to the number of intervals. To understand how well this approximation holds for the range of photons chosen, we estimated the number of lost photons when using these detectors.








For 10 $\mu$s pulses and a dead time of 40 ns, the number of measurement intervals is $N= 10000/40 = 250$. One or more photons are lost when they arrive in the same measurement interval.

\begin{figure}[ht]
    \centering
    \includegraphics[width=0.35\textwidth]{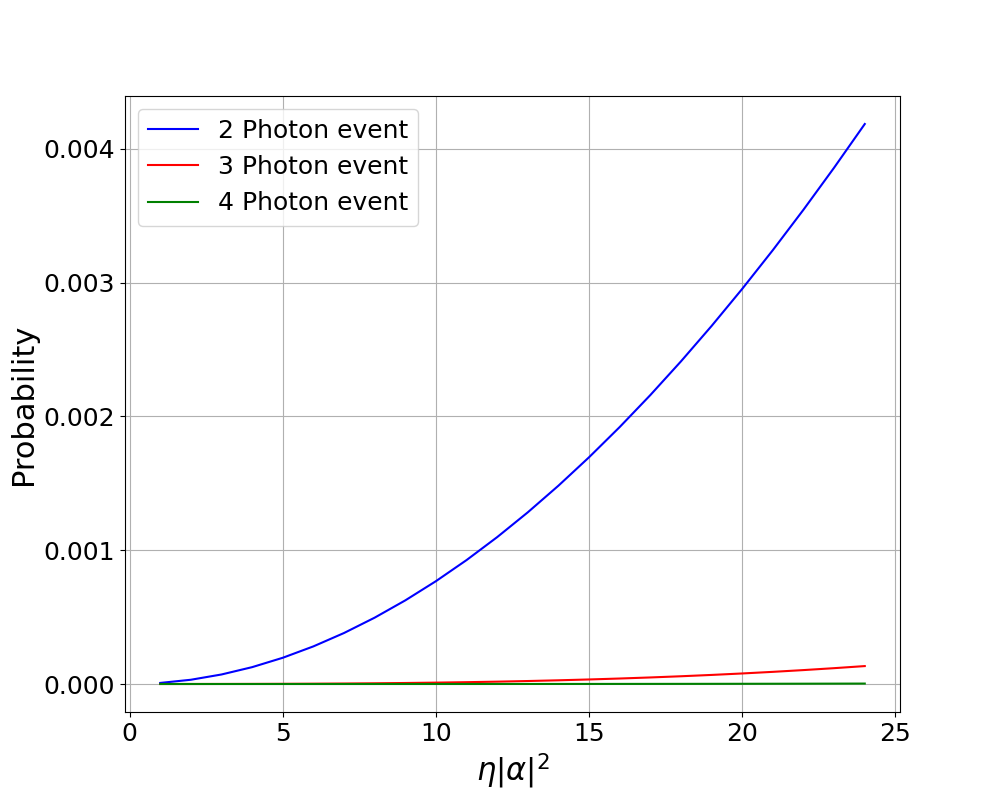} 
    \caption{Probability of n number of photons arriving to the detector in a single time interval as a function of the average number of photons detected.}
    \label{fig:event_prob}
\end{figure}

Fig \ref{fig:event_prob} shows the probability of detecting n photons within the same measurement interval, given that the photons are uniformly distributed across the pulse. The probability of observing this scenario is low when the average number of photons detected is small. However, given that each global measurement consists of 250 intervals, it is important to consider the cumulative likelihood of such events occurring over the entire measurement period.

The average number of photons lost $|\alpha|^2_{lost}$ can be estimated by:

\begin{equation}
    |\alpha|^2_{lost} = \sum_{n=1}^{\infty} n P(n||\alpha|^2),
\end{equation}

where $P(n||\alpha|^2)$ is the probability of losing n photons given $|\alpha|^2$ photons received. For this analysis we only consider events where in total, one, two or three photons are lost, $n\leq 3$.

\begin{figure}[ht]
    \centering
    \includegraphics[width=0.35\textwidth]{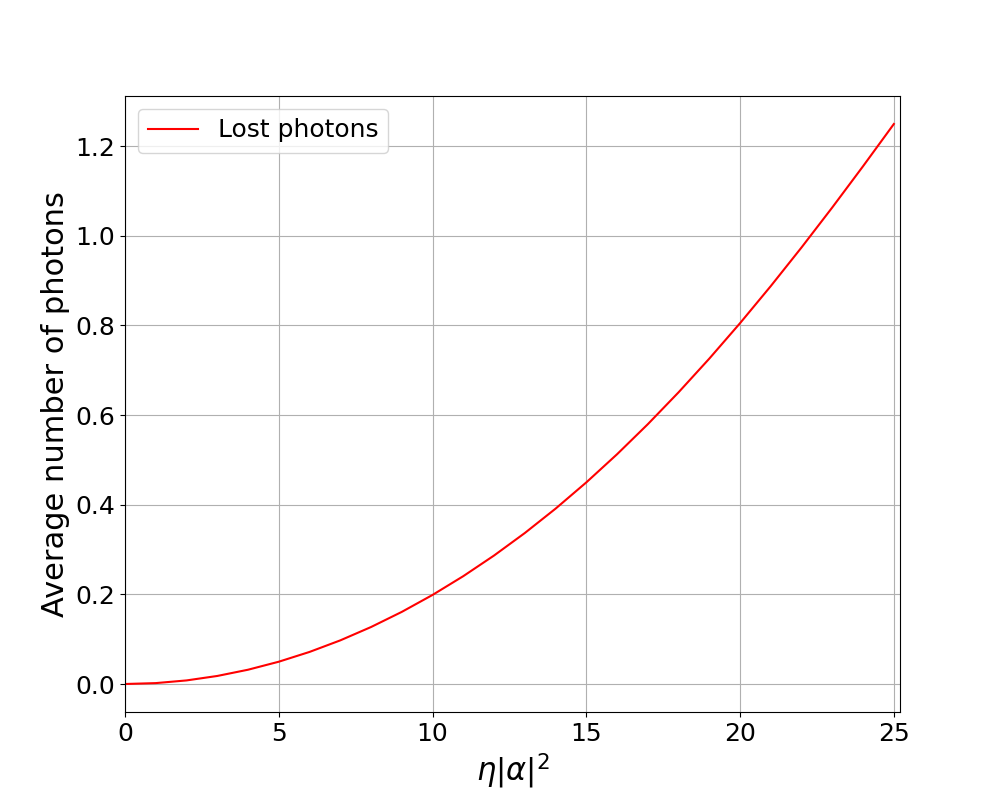} 
    \caption{Average number of photons lost as a function of the average number of photons received by the detector.}
    \label{fig:lost_photons}
\end{figure}

From Fig \ref{fig:lost_photons}, it can be seen that for a low average number of photons received, the average number of lost photons is low and the approximation is reasonable. 

While this technique can be useful for approximating a photon counting detector, it limits the rate of communication as the size of each pulse must be large, decreasing the repetition rate of the source and increasing the amount of photon noise arriving at the detectors.



    \label{eq:number of photons}




\section{Daylight Quantum Communication}
\label{appendix: background}

While most free-space QKD implementations can only be performed at nighttime, various works achieve a positive secret key rate during daytime \cite{avesani2021full, liao2017long, ko2018experimental}. This is done by filtering the background noise due to sunlight using spectral filters, spatial filters (by coupling the signal to single-mode fibers), or changing the operating wavelength among other techniques.

As seen in \cite{er2005background}, the background number of photons can be estimated using Eq. \ref{eq:background} to calculate the power received by the beam expander, $P_b$.

\begin{equation}
    P_b = H_b \times \Omega_{fov} \times A_{rec} \times B_{filter},
    \label{eq:background}
\end{equation}

where $H_b$ is the brightness of the sky background, $A_{rec}$ is the receiver's aperture, $\Omega_{fov}$ is the field of view, and $B_{filter}$ is the filter bandwidth. 

From Eq. \ref{eq:background} it can be seen that using a narrowband filter or decreasing the gating time of the detector can decrease the average number of photons arriving at the detector. However, a too-narrow filter can lead to wavelength stabilization problems and a too-short pulse can make the synchronization difficult. Reducing the receiver's aperture and the field of view can also be disadvantageous as it can increase the pointing requirements and reduce the amount of signal received. Table \ref{tab:noise_photons} shows the average photons per pulse for different weather conditions of a choice of these parameters. 

\begin{table}[h!]
\centering
\begin{tabular}{|c|c|c|c|}
\hline
\multirow{2}{*}{Conditions} & Relative & Typical Brightness & Photons \\
                            & Brightness & (W m$^{-2}$ Sr $\mu$m$^{-1}$) & per Pulse ($\Delta)$ \\ \hline
\makecell{Cloudy \\ Daytime}      & 1.0                          & 150                                                & 7.4                        \\ \hline
\makecell{Hazy \\Daytime }       & $10^{-1}$                    & 15                                                 & $7.4 \times 10^{-1}$       \\ \hline
\makecell{Clear\\ Daytime}       & $10^{-2}$                    & 1.5                                                & $7.4 \times 10^{-2}$       \\ \hline
\makecell{Full Moon \\Clear Night} & $10^{-5}$                  & $1.5 \times 10^{-3}$                               & $7.4 \times 10^{-5}$       \\ \hline
\makecell{New Moon\\ Clear Night} & $10^{-6}$                   & $1.5 \times 10^{-4}$                               & $7.4 \times 10^{-6}$       \\ \hline
\makecell{Moonless \\Clear Night} & $10^{-7}$                   & $1.5 \times 10^{-5}$                               & $7.4 \times 10^{-7}$       \\ \hline
\end{tabular}
\caption{Comparison of the number of noise photons received per pulse under different weather conditions. The value of photons per pulse was calculated for 100 $\mu$rad of divergence half-angle, 0.2 nm spectral filter, 850 nm wavelength, and 3 ns of gating time. Adapted from \cite{er2005background}.}
\label{tab:noise_photons}
\end{table}

As seen in table \ref{table:background}, this work, without specific techniques to reduce background noise, already has a photon noise one order of magnitude higher than daylight QKD experiments and still achieves secure communication. 

\begin{figure}[ht]
    \centering
    \includegraphics[width=0.35\textwidth]{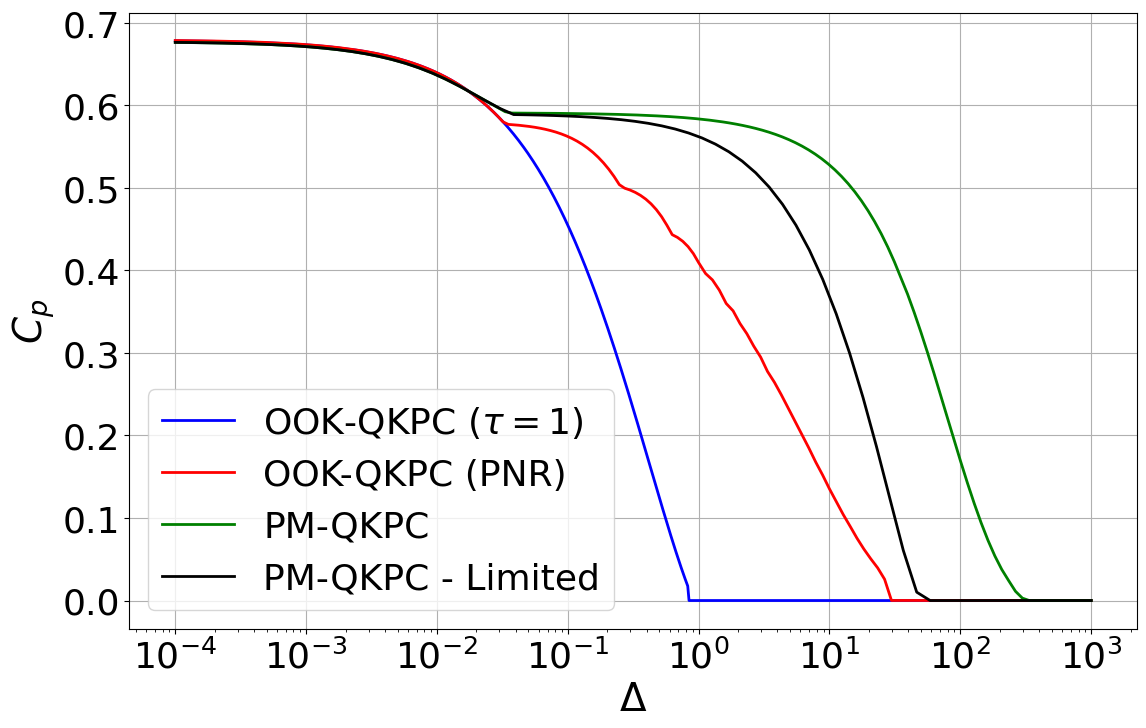} 
    \caption{Plot of the maximum private capacity as a function of the photon noise, $\Delta$, for $\gamma=0.1$ The different colors correspond to OOK-QKPC without PNR measurements (blue), OOK-QKPC with PNR measurements (red), PM-QKPC (green) and PM-QKPC with limited polarization angle and average number of photons (black).}
    \label{fig:2d_fixed_gamma_app}
\end{figure}

From Fig. \ref{fig:2d_fixed_gamma_app}, PM-QKPC allows secret communication with high private capacity even for a background number of photons 3-4 orders of magnitude higher than in the daylight QKD experiments. This allows for a reduction of the complexity of the experimental setups used.

\begin{table}[ht]
\centering
\begin{tabular}{|c|c|c|c|}
\hline

& \textbf{This work} & \cite{avesani2021full} & \cite{ko2018experimental} \\
\hline
Source frequency (Hz) & $50 \cdot 10^{3}$ & $50 \cdot 10^{6}$ & $100 \cdot10^{6}$ \\
Photon noise (Hz) &   $1.5 \cdot 10^{3}$  & $240$ & $578 \cdot 10^{3}$\\
Photon noise (per pulse) &  $0.03$  & $0.0000048$ & $0.0058$\\
\hline

\end{tabular}
\caption{Photon noise estimation for different daylight quantum communication experiments.}
\label{table:background}
\end{table}





\section{Unambiguous State Discrimination}
\label{appendix: USD}

We extend our analysis by considering the case in which Bob implements unambiguous state discrimination (USD) rather than a minimum-error strategy. In USD, Bob’s measurement sometimes yields an inconclusive outcome (an erasure) but, whenever a conclusive result is obtained, it is guaranteed to be correct. 

This behavior is modeled by a binary erasure channel, where the probability of an inconclusive, $p_i$ (or “erasure”) result is determined by the overlap squared between the transmitted states, Eq. \ref{Eq: Polarization states}.

The mutual information between Alice and Bob is then given by the channel capacity of a binary erasure channel:

\begin{equation}
    I_B = 1 - e^{-\eta|\alpha|^2(1 - \cos{\theta})}.
\end{equation}

\begin{figure}[h!]
    \centering
    \includegraphics[width=0.38\textwidth]{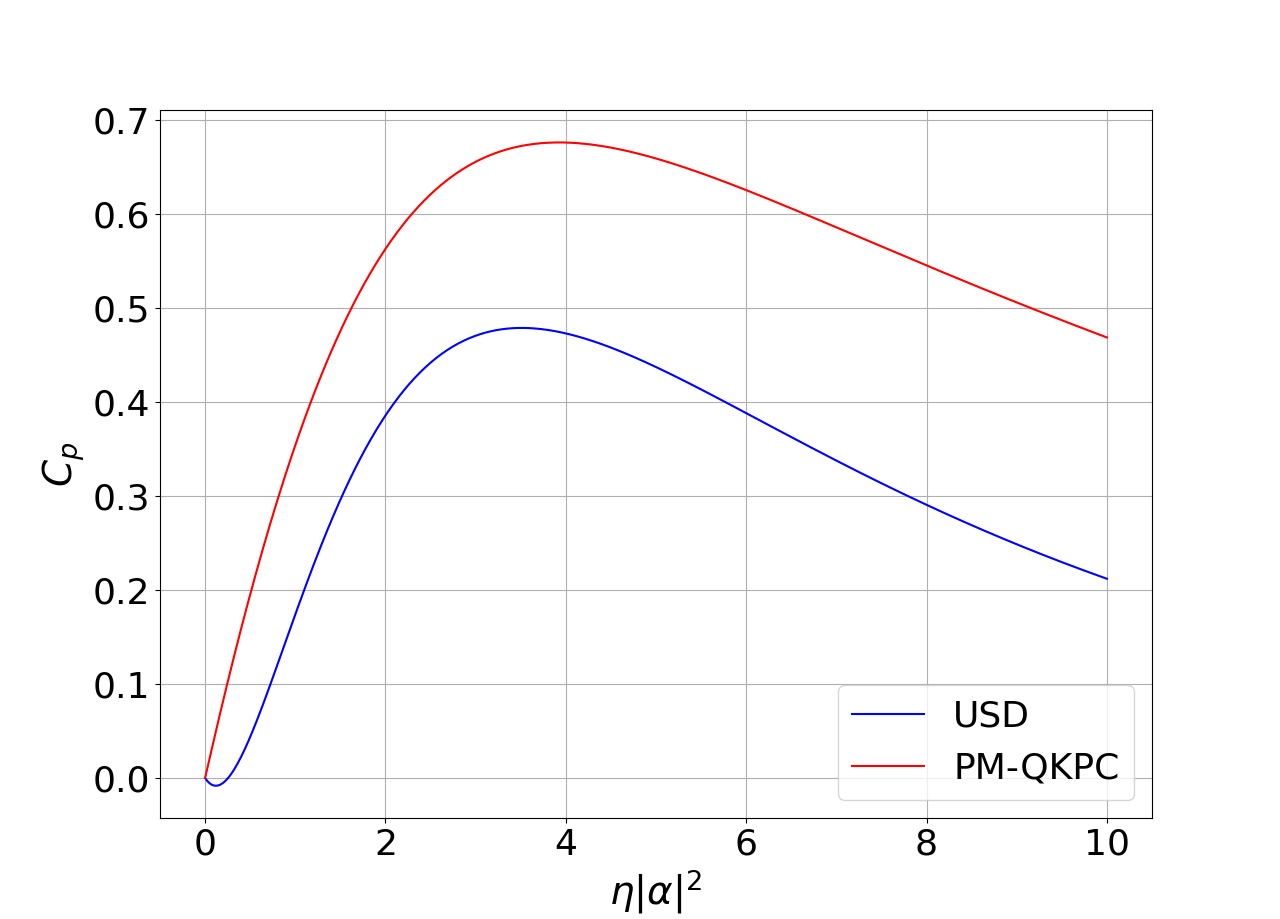} 
    \caption{Private capacity as a function of the received number of photons. In red is the PM-QKPC private capacity and in blue is the private capacity using USD. Values taken for $\gamma=0.1$.}
    \label{fig:USD}
\end{figure}

Eve is assumed to be able to choose between the USD or minimum-error strategy. 

\begin{equation}
    C_p(\gamma) = \text{max}_{|\alpha|^2,\tau} \left[I_B - I_E\right].
    \label{eq: Private Capacity_USD}
\end{equation}

Using Eq. \ref{eq: Private Capacity_USD}, we can estimate the private capacity for any $\gamma$.  We find that it is always more advantageous for Eve to choose minimum error discrimination. Figure \ref{fig:USD} shows the private capacity for when Bob uses USD or minimum error discrimination. It can be seen that the private capacity using USD is always lower than using the minimum-error strategy, meaning there is no advantage in using USD.

\end{document}